\documentclass[a4paper,11pt]{article}

\usepackage{pos}
\usepackage{mine}

\graphicspath{{figures/}}

\title{Theory pipeline for PDF fitting}
\ShortTitle{Theory pipeline}

\author[a]{Andrea Barontini}
\author*[a]{Alessandro Candido}
\author[a,b]{Juan M. Cruz-Martinez}
\author[a]{Felix Hekhorn}
\author[c]{Giacomo Magni}
\author[d]{Christopher Schwan}

\affiliation[a]{
  TIF Lab, Dipartimento di Fisica, Universit\`a degli Studi di Milano and INFN
  Sezione di Milano \\
  Via Celoria 16, 20133, Milano, Italy
}
\affiliation[b]{CERN, Theoretical Physics Department, CH-1211 Geneva 23, Switzerland}
\affiliation[c]{
  Department of Physics and Astronomy, Vrije Universiteit, Amsterdam, The Netherlands
  Nikhef Theory Group, Amsterdam, The Netherlands
}
\affiliation[d]{
  Universit\"at W\"urzburg, Institut f\"ur Theoretische Physik und
  Astrophysik, 97074 W\"urzburg, Germany
}

\emailAdd{andrea.barontini@mi.infn.it}
\emailAdd{alessandro.candido@mi.infn.it}
\emailAdd{juan.cruz.martinez@cern.ch}
\emailAdd{felix.hekhorn@mi.infn.it}
\emailAdd{christopher.schwan@physik.uni-wuerzburg.de}

\abstract{
  Fitting \pdf{}s requires the integration of a broad range of
  datasets, both from data and theory side, into a unique framework.
  While for data the integration mainly consists in the standardization of the
  data format, for the theory predictions there are multiple ingredients
  involved.
  Different providers are developed by separate groups for different processes,
  with a variety of inputs (runcards) and outputs (interpolation grids).
  Moreover, since processes are measured at different scales, \dglap evolution
  has to be provided for the \pdf candidate, or precomputed into the grids.
  We are working towards the automation of all these steps in a unique framework,
  that will be useful for any \pdf fitting groups, and possibly also for
  phenomenological studies.
}

\FullConference{%
  41st International Conference on High Energy physics - ICHEP2022\\
  6-13 July, 2022\\
  Bologna, Italy
}

\begin{document}
\maketitle

\section{Introduction}
\label{sec:intro}

Performing a \pdf fit requires to integrate several elements to be gathered
from many different sources: data from several experiments, ranging over
multiple decades and formats, and competitive theory predictions, coming
from different providers.
Finally, a fitting methodology has to be selected and engineered to implement
theory constraints, and to limit not physically motivated bias.

While data is a \textit{static} component in the fit, the theory predictions
depend on the candidate \pdf, since they are the mapping that connect the
unobserved \pdf space, to the observed data space.
During the fit, this map will be used a large number of times (at least once for
every minimization step), so it is paramount to have an efficient way to
evaluate it, otherwise it can become a serious bottleneck.

For this reason, a few interfaces to \pdf independent theory predictions have
already been implemented
\cite{Carli:2010rw,Britzger:2012bs,Britzger:2022lbf,Carrazza:2020gss}, and they
are used in different contexts.
They propose different file formats to store the output of a Monte Carlo
generator, splitting them by luminosity component, perturbative order, and
observables binning.
This output can be optimized as an \textit{interpolation grid}, leveraging the
fact that the \pdf itself is only defined over a finite set of points, and thus
including the interpolation basis in the factorized cross-section.
Essentially, this recast the partonic cross sections predictions as
a \textit{theory array}, for which the Mellin convolution is replaced by a
linear algebra contraction over a single or multiple \pdf set.
This idea can be broadened to apply to any factorized function, describing the
structure of an external hadron (both incoming and outgoing).

However, this picture does not exhaust the needs of a \pdf fit (or any other
hadronic one), because, while the \pdf dependence on flavor and $x$ value is
folded on the grid, that on factorization scale has to be fixed to the process
dependent value.
This dependence is not fitted, since it is only determined by perturbative
\qcd.
In order to obtain it, it is required to solve the \dglap equation with the
border condition provided by the fit.
But being \dglap a set of integro-differential equations \textit{linear} in the
\pdf, this can be converted in the application of a suitable \textit{evolution
operator}, solving the same equation.
Since the evolution operator can also be computed ahead of time, it is possible
to combine the two ingredients (the operator and the grid) in a single fast
array interface, that will directly produce the required theory predictions
once contracted on the \pdf candidate.
Thus, the map from \pdf space to data space discussed above, is reduced to a
linear algebra product (or more than one, when multiple hadrons are involved).
During this operation, there is no loss of generality, since the interpolation
basis used for the conversion of the analytic convolutions is already present
in many \pdf applications due to their non-perturbative nature.
Such an interface is called a \enquote{Fast Kernel table} (shortened to
\textit{FK table}) in the context of the \nnpdf collaboration.

To produce the \textit{\fktab{}s} an evolution operator provider is required,
and needs to be interfaced with the grids. This was originally done in \nnpdf
by an internal tool (\fkgen), and then systematized in the \apfelgrid
\cite{Bertone:2016lga} package (leveraging the \apfel~\cite{Bertone:2013vaa}
evolution library), later reworked once more taking the name of \apfelcomb
\cite{APFELcomb}.

An array interface is extremely useful, since it allows to treat the theory map
in the context of many software frameworks, just relying on the data structure
of an array. Especially relevant for machine
learning software frameworks, but not limited to them, e.g.\ it allows to
create a Bayesian inference-based methodology, without the need of the
treatment of further complex functions.

\vspace*{-5pt}
\section{Architecture}
\label{sec:arch}

As it has been explained in the previous section, a theory map, i.e.\ an
\fktab, is made of two main components: a \pdf independent interpolation
\textit{grid} and an \textit{evolution operator}.

For the second one, we just need a single provider, able to compute the \dglap
solving operator for a variety of theory settings (corresponding to different
\pdf fits, e.g.\ \nlo and \nnlo \qcd evolution), able to perform the
operator computation as efficient as possible, and to smoothly interface with
the grid for convolution.
Multiple evolution codes are already available for the purpose
\cite{Vogt:2004ns,Salam:2008sz,Botje:2010ay,Bertone:2013vaa,Bertone:2017gds},
but when the authors decided to begin the full rework of the architecture, it
has been clear that a dedicated tool, with this exact goal in mind, would have
been an ideal solution.
For this reason, the software package \eko \cite{Candido:2022tld} has been
created, providing a framework to solve \dglap equations in Mellin space,
similarly to \cite{Vogt:2004ns}, but generating an output in $x$-space, i.e.\
the one usually employed by the fit.
Moreover, while not being the first tool to provide evolution operators
\cite{Bertone:2013vaa,Salam:2008sz}, it produces them as the default output,
optimizing the process as much as possible, storing them in a dedicated file
format, designed to be easily available from different programming languages.

\eko is very different from \apfel, the tool on which the NNPDF framework has
until now relied. For instance \apfelgrid (then \apfelcomb), the tool which
generates \apfel-based \fktab, introduces an explicit dependency on \apfel
itself (and thus its internals).
\eko instead not only exposes a restricted public API (making all the dependent
projects decoupled from its very internals), but the dependency is not required
at all to consume the \eko output, consisting of float arrays stored in a very
common \href{https://en.wikipedia.org/wiki/Tar\_(computing)}{\texttt{tar}}
archive, and standard \href{https://yaml.org/}{\texttt{YAML}} metadata.
On the other side, the observable grids have to be produced by different
generators, in order to cover the full variety of available processes.
For this reason, we need an interface to them, with the following targets:
standardizing the output and making it reproducible.

The solution we propose is thus based on the concept of interpolation grid, and
specifically on \pineappl as an interface.
In particular, \pineappl exposes APIs to different languages: it is natively
written in Rust, but has an API to C/C++, that can be consumed also by a
Fortran application (examples provided for all of them), and a Python API,
mostly dedicated to scripting and integration with the rest of the pipeline,
but there are providers (essentially \yadism
\cite{candido_alessandro_2022_6285149}, used for DIS at NNLO) already using it
to fill grids.

\begin{figure}
  \centering
  \includegraphics[width=0.9\textwidth]{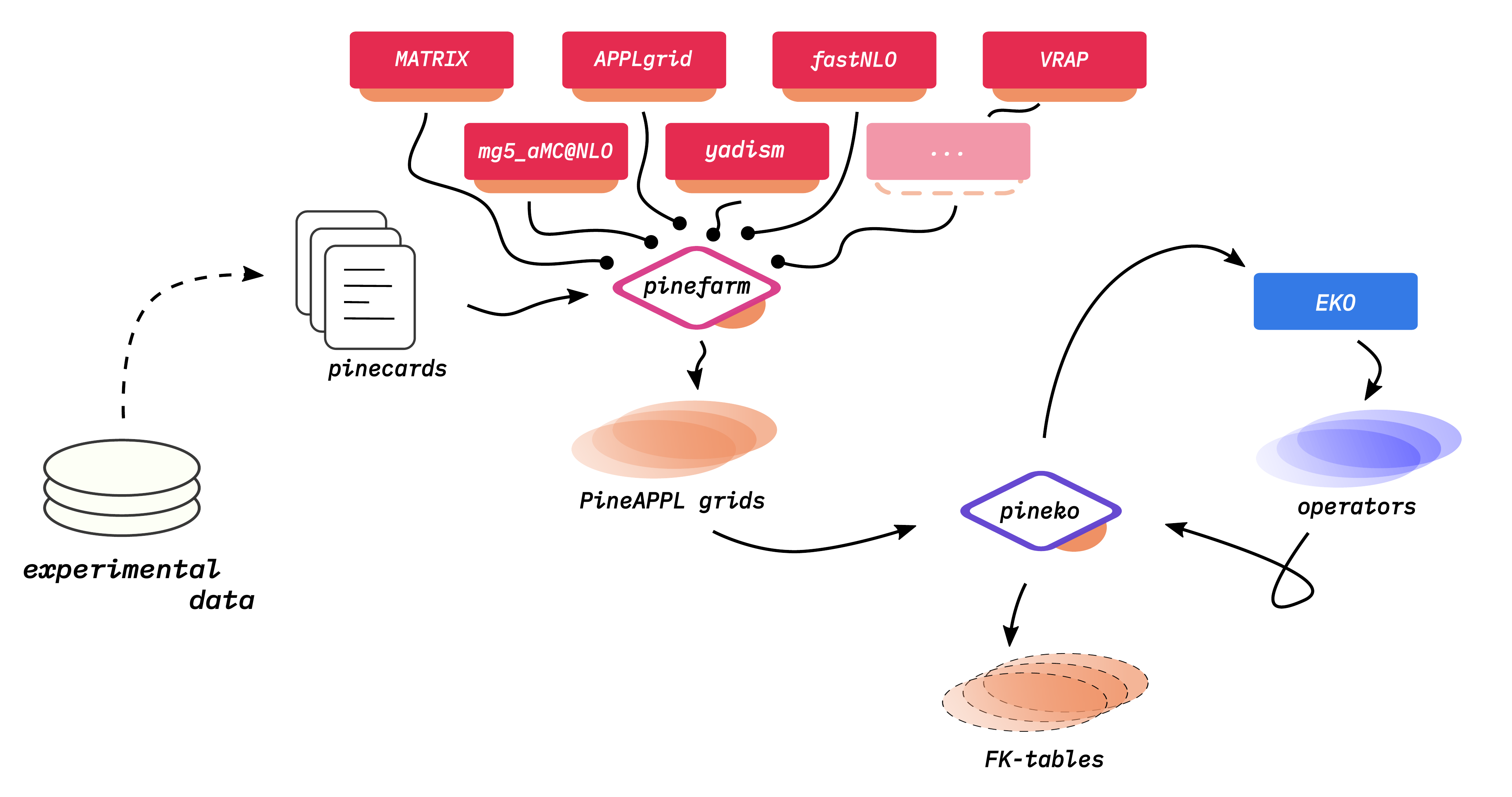}
  \caption{
    Updated version of the flow diagram already appeared in
    \cite{Amoroso:2022eow}, showing the overall pipeline architecture.
    Arrows in the picture indicate the flow of information (together with the
    execution order), and the orange insets on other elements indicate an
    interface to \pineappl (notice \eko not having it).
    In particular, magenta blocks above \pinefarm are the providers
    \cite{Grazzini:2017mhc,Frederix:2018nkq,Carli:2010rw,candido_alessandro_2022_6285149,Britzger:2012bs,Anastasiou:2003ds}.
  }
  \label{fig:pineline}
  \vspace*{-5pt}
\end{figure}

Since different generators require different inputs, we are trying to
standardize them into a common format for which other cards can be generated,
called \textit{pinecard}.
This is still work in progress, nevertheless, it is useful to speak of
pinecards, since they are used as inputs for \pinefarm, that is the unique
Python package working as a front-end for the various generators.
Essentially, each generator needs dedicated code to run, but this interface has
to be written once, and then is part of \pinefarm, standardizing the input for
that generator, and part of the input across all of them (e.g.\ metadata, like
references and observable details, or theory parameters).
In \cref{fig:pineline} we summarize our architecture: the generators are
directly interfaced with the \pineappl library, and the output is thus
standardized to an interpolation grid (for one or two colliding hadrons), the
input instead consists of a \textit{pinecard}.

Once the grid is available, \pineko (a package dedicated to the final
construction of \fktab{}s) can extract the details of the operator needed for
the \fktab generation from the grid, generate the \eko input, and then combine
the grid and the operator into the final \fktab.

All the components of the pipeline are open source and the code is available in
the NNPDF GitHub organization:
\begin{itemize}
  \setlength\itemsep{2pt}
  \item \pineappl: \url{https://github.com/NNPDF/pineappl}
  \item \eko: \url{https://github.com/NNPDF/eko}
  \item \pineko: \url{https://github.com/NNPDF/pineko}
  \item \pinefarm: \url{https://github.com/NNPDF/runcards} (including the
    relevant pinecards, required for \nnpdf fits)
\end{itemize}

The set of tools does not depend
on the \nnpdf fitting methodology and can be used in general for
any hadronic function fitting\footnote{
  Generalization of \pineappl to support fragmentation functions and polarized
  \pdf{}s is work in progress.
}.

\section{Applications}
\label{sec:appl}

Components have applications on their own, and part of them have already been
used (or are being used) to support other works.
Even though here it appears incidental, this is an important design feature: we
are building a framework, not just a pipeline application. The
various components should be focused on dedicated tasks and easy to integrate
in different architecture (or, more realistically, stand-alone projects), for
similar but different goals.

A first example is the study on evidence
for an intrinsic charm component in the proton \cite{Ball:2022qks}, based on
the \nnpdf 4.0 \pdf set, latest release of the \nnpdf family, and \eko
\cite{Candido:2022tld}, the evolution code described in the previous
\cref{sec:arch}.
The role of \eko has been to unfold the intrinsic component from the so-called
fitted charm, in the 4 flavor number scheme (default scheme at fitting scale
for \nnpdf), by backward evolving with \dglap equation in a 3 flavor number
scheme \pdf set at a lower scale.
On top of the required backward evolution, and the proper treatment of
intrinsic components,
\eko implemented the \nnnlo matching conditions
between the 4 and 3 flavor schemes, that have been relevant to estimate the
perturbative stability of the result obtained.

Another application is the study of the forward backward asymmetry in the
Drell--Yan process with a high cut in the invariant mass of the
lepton pair \cite{Ball:2022qtp}.
In particular, the work focuses on the comparison between results obtained with
the \nnpdf 4.0 \pdf set and other contemporary \pdf sets from different
collaborations. We find that a certain shape in the high cut setting is
related to the specific shape of the \pdf{}s in the large-$x$ extrapolation
region, and so very sensitive to the possible bias of extending behaviors
typical of the central data region.
In this context, it has been crucial to have \pineappl \cite{Carrazza:2020gss}
\cite{christopher_schwan_2022_7145377} grids pre-computed to reproduce the
results, iterating on the \pdf set to investigate different features of the
\pdf, and trying to trace back the distribution behavior to \pdf features.

Finally, a study of the low energy neutrino structure functions is ongoing,
where the low $Q^2$ experimental data is reconciled to the known
perturbative calculation at higher energies, based on the \pdf{}s.
Here, we use \yadism, a general inclusive DIS provider
interfaced with \pineappl, to produce perturbative \qcd calculation for the
structure functions that get matched to experimental data.

\section{Conclusions}
\label{sec:concl}

In this proceeding we presented the new theory predictions framework and described
its main application in the context of \pdf fitting.
The main features are the standardization, maintainability, modularity, and
reproducibility. We aim for a good support for multiple users and external
contributions. Finally, we discussed some early stand-alone applications for the
individual components introduced in the framework.

\acknowledgments
AB, AC, JCM, and FH are supported by the European Research Council under the
European Union's Horizon 2020 research and innovation Programme (grant
agreement number 740006).
C.S.\ is supported by the German Research Foundation (DFG) under reference
number DE 623/6-2.

\bibliographystyle{JHEP}
\bibliography{blbl}

\providecommand{\href}[2]{#2}\begingroup\raggedright\begin{thebibliography}{10}

\bibitem{Carli:2010rw}
T.~Carli, D.~Clements, A.~Cooper-Sarkar, C.~Gwenlan, G.~P. Salam, F.~Siegert
  et~al., \emph{{A posteriori inclusion of parton density functions in NLO QCD
  final-state calculations at hadron colliders: The APPLGRID Project}},
  \href{https://doi.org/10.1140/epjc/s10052-010-1255-0}{\emph{Eur. Phys. J. C}
  {\bfseries 66} (2010) 503} [\href{https://arxiv.org/abs/0911.2985}{{\ttfamily
  0911.2985}}].

\bibitem{Britzger:2012bs}
{\scshape fastNLO} collaboration, \emph{{New features in version 2 of the
  fastNLO project}},  in \emph{{20th International Workshop on Deep-Inelastic
  Scattering and Related Subjects}}, pp.~217--221, 2012,
  \href{https://doi.org/10.3204/DESY-PROC-2012-02/165}{DOI}
  [\href{https://arxiv.org/abs/1208.3641}{{\ttfamily 1208.3641}}].

\bibitem{Britzger:2022lbf}
D.~Britzger et~al., \emph{{NNLO interpolation grids for jet production at the
  LHC}},  \href{https://arxiv.org/abs/2207.13735}{{\ttfamily 2207.13735}}.

\bibitem{Carrazza:2020gss}
S.~Carrazza, E.~R. Nocera, C.~Schwan and M.~Zaro, \emph{{PineAPPL: combining EW
  and QCD corrections for fast evaluation of LHC processes}},
  \href{https://doi.org/10.1007/JHEP12(2020)108}{\emph{JHEP} {\bfseries 12}
  (2020) 108} [\href{https://arxiv.org/abs/2008.12789}{{\ttfamily
  2008.12789}}].

\bibitem{Bertone:2016lga}
V.~Bertone, S.~Carrazza and N.~P. Hartland, \emph{{APFELgrid: a high
  performance tool for parton density determinations}},
  \href{https://doi.org/10.1016/j.cpc.2016.10.006}{\emph{Comput. Phys. Commun.}
  {\bfseries 212} (2017) 205}
  [\href{https://arxiv.org/abs/1605.02070}{{\ttfamily 1605.02070}}].

\bibitem{Bertone:2013vaa}
V.~Bertone, S.~Carrazza and J.~Rojo, \emph{{APFEL: A PDF Evolution Library with
  QED corrections}},
  \href{https://doi.org/10.1016/j.cpc.2014.03.007}{\emph{Comput. Phys. Commun.}
  {\bfseries 185} (2014) 1647}
  [\href{https://arxiv.org/abs/1310.1394}{{\ttfamily 1310.1394}}].

\bibitem{APFELcomb}
V.~Bertone, S.~Carrazza and N.~Hartland,
  ``\url{https://github.com/NNPDF/apfelcomb}.''.

\bibitem{Vogt:2004ns}
A.~Vogt, \emph{{Efficient evolution of unpolarized and polarized parton
  distributions with QCD-PEGASUS}},
  \href{https://doi.org/10.1016/j.cpc.2005.03.103}{\emph{Comput. Phys. Commun.}
  {\bfseries 170} (2005) 65}
  [\href{https://arxiv.org/abs/hep-ph/0408244}{{\ttfamily hep-ph/0408244}}].

\bibitem{Salam:2008sz}
G.~Salam and J.~Rojo, \emph{{The HOPPET NNLO parton evolution package}},  in
  \emph{{16th International Workshop on Deep Inelastic Scattering and Related
  Subjects}}, p.~42, 7, 2008, \href{https://doi.org/10.3360/dis.2008.42}{DOI}
  [\href{https://arxiv.org/abs/0807.0198}{{\ttfamily 0807.0198}}].

\bibitem{Botje:2010ay}
M.~Botje, \emph{{QCDNUM: Fast QCD Evolution and Convolution}},
  \href{https://doi.org/10.1016/j.cpc.2010.10.020}{\emph{Comput. Phys. Commun.}
  {\bfseries 182} (2011) 490}
  [\href{https://arxiv.org/abs/1005.1481}{{\ttfamily 1005.1481}}].

\bibitem{Bertone:2017gds}
V.~Bertone, \emph{{APFEL++: A new PDF evolution library in C++}},
  \href{https://doi.org/10.22323/1.297.0201}{\emph{PoS} {\bfseries DIS2017}
  (2018) 201} [\href{https://arxiv.org/abs/1708.00911}{{\ttfamily
  1708.00911}}].

\bibitem{Candido:2022tld}
A.~Candido, F.~Hekhorn and G.~Magni, \emph{{EKO: Evolution Kernel Operators}},
  \href{https://arxiv.org/abs/2202.02338}{{\ttfamily 2202.02338}}.

\bibitem{candido_alessandro_2022_6285149}
A.~Candido, F.~Hekhorn and G.~Magni, \texttt{N3PDF/yadism: FONLL-B}
  \texttt{v0.11.0} (Feb., 2022),
  \href{https://doi.org/10.5281/zenodo.6285149}{doi:10.5281/zenodo.6285149}.

\bibitem{Amoroso:2022eow}
S.~Amoroso et~al., \emph{{Snowmass 2021 whitepaper: Proton structure at the
  precision frontier}},  \href{https://arxiv.org/abs/2203.13923}{{\ttfamily
  2203.13923}}.

\bibitem{Grazzini:2017mhc}
M.~Grazzini, S.~Kallweit and M.~Wiesemann, \emph{{Fully differential NNLO
  computations with MATRIX}},
  \href{https://doi.org/10.1140/epjc/s10052-018-5771-7}{\emph{Eur. Phys. J. C}
  {\bfseries 78} (2018) 537}
  [\href{https://arxiv.org/abs/1711.06631}{{\ttfamily 1711.06631}}].

\bibitem{Frederix:2018nkq}
R.~Frederix, S.~Frixione, V.~Hirschi, D.~Pagani, H.~S. Shao and M.~Zaro,
  \emph{{The automation of next-to-leading order electroweak calculations}},
  \href{https://doi.org/10.1007/JHEP07(2018)185}{\emph{JHEP} {\bfseries 07}
  (2018) 185} [\href{https://arxiv.org/abs/1804.10017}{{\ttfamily
  1804.10017}}].

\bibitem{Anastasiou:2003ds}
C.~Anastasiou, L.~J. Dixon, K.~Melnikov and F.~Petriello, \emph{{High precision
  QCD at hadron colliders: Electroweak gauge boson rapidity distributions at
  NNLO}}, \href{https://doi.org/10.1103/PhysRevD.69.094008}{\emph{Phys. Rev. D}
  {\bfseries 69} (2004) 094008}
  [\href{https://arxiv.org/abs/hep-ph/0312266}{{\ttfamily hep-ph/0312266}}].

\bibitem{Ball:2022qks}
{\scshape NNPDF} collaboration, \emph{{Evidence for intrinsic charm quarks in
  the proton}}, \href{https://doi.org/10.1038/s41586-022-04998-2}{\emph{Nature}
  {\bfseries 608} (2022) 483}
  [\href{https://arxiv.org/abs/2208.08372}{{\ttfamily 2208.08372}}].

\bibitem{Ball:2022qtp}
R.~D. Ball, A.~Candido, S.~Forte, F.~Hekhorn, E.~R. Nocera, J.~Rojo et~al.,
  \emph{{Parton Distributions and New Physics Searches: the Drell-Yan
  Forward-Backward Asymmetry as a Case Study}},
  \href{https://arxiv.org/abs/2209.08115}{{\ttfamily 2209.08115}}.

\bibitem{christopher_schwan_2022_7145377}
C.~Schwan, A.~Candido, F.~Hekhorn and S.~Carrazza, \texttt{NNPDF/pineappl:
  v0.5.7} \texttt{v0.5.7} (Oct., 2022),
  \href{https://doi.org/10.5281/zenodo.7145377}{doi:10.5281/zenodo.7145377}.

\end{thebibliography}\endgroup

\listoffixmes

\end{document}